\documentclass[sigconf,nonacm]{acmart}
\usepackage{courier}

\makeatletter
\def\@ACM@bibliography@startnumber{1}
\makeatother
\title{Roadmap to Quantum Aesthetics}

\author{Ivan C. H. Liu}
\orcid{0000-0002-9226-9765}
\affiliation{
  \institution{Future Narratives Lab\\Institute of Applied Arts\\National Yang Ming Chiao Tung University,}
  \city{Hsinchu}
  \country{Taiwan (R.O.C)}
}
\email{ivanliu@nycu.edu.tw}
\author{Hsiao-Yuan Chen}
\affiliation{
  \institution{Future Narratives Lab\\Institute of Applied Arts\\National Yang Ming Chiao Tung University,}
  \city{Hsinchu}
  \country{Taiwan (R.O.C)}
}

\begin{document}

\begin{abstract}
Quantum mechanics occupies a central position in contemporary science while remaining largely inaccessible to direct sensory experience. This paper proposes a roadmap to quantum aesthetics that examines how quantum concepts become aesthetic phenomena through artistic mediation rather than direct representation. Two complementary and orthogonal approaches are articulated. The first, a pioneering \emph{top-down} approach, employs text-prompt-based generative AI to probe quantum aesthetics as a collective cultural construct embedded in large-scale training data. By systematically modulating the linguistic weight of the term ``quantum,'' generative models are used as experimental environments to reveal how quantum imaginaries circulate within contemporary visual culture. The second, a \emph{bottom-up} approach, derives aesthetic form directly from quantum-mechanical structures through the visualization of quantum-generated data, exemplified here by hydrogen atomic orbitals calculated from the Schr\"{o}dinger equation. These approaches are framed not as competing methods but as intersecting paths within a navigable field of artistic research. They position quantum aesthetics as an emergent field of artistic research shaped by cultural imagination, computational mediation, and physical law, opening new directions for artistic practice and pedagogy at the intersection of art, data, artificial intelligence and quantum science.
\end{abstract} %%%%%%%%%
\maketitle

\section{Introduction}

Quantum mechanics occupies a distinctive position in both scientific inquiry and contemporary culture. While it provides the theoretical foundation for modern physics and emerging quantum technologies, its central concepts, such as superposition, uncertainty, entanglement, and nonlocality, remain difficult to reconcile with everyday perception. This gap between mathematical formalism and sensory intuition has made quantum theory a recurring reference in artistic discourse. Across media art and related practices, quantum ideas have been used to address indeterminacy, multiplicity, and the limits of classical representation. Yet the question of what constitutes a quantum aesthetic, and how such an aesthetic can be articulated beyond metaphor, remains open.

In recent artistic practice, engagement with quantum mechanics has increasingly shifted from symbolic reference toward operational and generative use. Rather than illustrating quantum concepts, artists incorporate quantum equations, simulations, data structures, or computational processes directly into the compositional workflow. For example, works by Putnam and coworkers use hydrogen atom wavefunctions derived from the Schr\"{o}dinger equation as structural inputs for audiovisual composition, treating electron probability densities as formal constraints that guide aesthetic outcomes \cite{putnam2015hydrogen}. Subsequent projects extend this approach into immersive and performative contexts, where quantum data are spatialized and navigated through large-scale simulations, emphasizing scale, temporality, and embodied perception \cite{kuchera2022complex}.

Theoretical perspectives have developed alongside these practices. Thomas frames uncertainty and entanglement as creative conditions that destabilize the classical separation between observer and observed, aligning quantum indeterminacy with open-ended artistic processes \cite{thomas2018quantum}. Related technological explorations by Heaney (2019) appear in works that engage quantum computation \cite{heaney2019quantum}. Heaney’s practice employs quantum computation as an aesthetic condition rather than a technical novelty.

Institutional initiatives also reflect this operational turn. An example is LAS Art Foundation's \emph{Sensing Quantum} project, an interdisciplinary collaboration between artists and quantum scientists, translating quantum processes into sensory and experiential forms, emphasizing perception and speculation rather than metaphor or didactic representation, and positioning artistic research as a mediator between scientific abstraction and embodied experience \cite{las2020sensing}.

While these works engage quantum mechanics through physical theory, simulation, or hardware, the present study introduces a distinct and previously underexplored approach: the use of generative AI to probe quantum aesthetics as a collective cultural construct. This work pioneers the systematic use of text-prompt-based generative AI to extract, synthesize, and modulate aggregated visual associations embedded in large-scale training datasets. Rather than treating generative AI as a tool for stylistic production alone, we position it as an epistemic instrument capable of revealing how ``quantum'' is already encoded within collective aesthetic experience. In this framework, generative AI enables speculative access to cultural imaginaries of the quantum that precede, exceed, or diverge from scientific formalism.

Building on this evolving landscape, the present paper proposes a roadmap to quantum aesthetics structured around two complementary methodological approaches: a \emph{top-down} approach and a \emph{bottom-up} approach. The roadmap does not aim to unify these approaches under a single definition; instead, it frames them as distinct yet interrelated strategies for examining how quantum concepts become aesthetic phenomena through different mediating systems.

The \emph{top-down} approach examines quantum aesthetics as they emerge within contemporary visual culture through text-prompt-based generative AI systems. Diffusion-based image synthesis models are trained on large-scale datasets drawn from digital culture, encoding collective visual conventions and stylistic associations. Within this context, quantum functions as a culturally charged linguistic modifier, activating associations with abstraction, energy, fluidity, luminosity, and speculative futurity. By systematically employing prompts such as ``quantum” and modulating their relative weights, generative AI is used as an experimental environment to probe how quantum aesthetics are culturally constructed and algorithmically negotiated. The resulting images do not claim correspondence with physical quantum phenomena; rather, they reveal how quantum ideas circulate and are aestheticized within collective imagination mediated by computational systems.

The \emph{bottom-up} approach, by contrast, begins with quantum mechanics itself. Here, aesthetic form is derived directly from quantum-mechanical equations and datasets. This paper focuses on visualizations of hydrogen atomic orbitals calculated from the Schr\"{o}dinger equation, producing probability density distributions governed by strict mathematical constraints. These forms exhibit symmetry, nodal structure, and increasing complexity across quantum states. When examined through theories of information aesthetics and contemporary visualization discourse, such images can be understood as aesthetic artifacts rather than neutral scientific illustrations, with aesthetic qualities arising from the structured organization of information and physical constraint \cite{lau2011information}. 

Taken together, the two approaches interrogate quantum aesthetics as a distributed phenomenon emerging through the interaction of cultural imagination, computational mediation, and physical law. Rather than defining a singular quantum style, the roadmap positions quantum aesthetics as an open field of artistic research in which scientific theory, generative systems, and aesthetic experience co-evolve. Moreover, these two approaches represent two orthogonal modes of creative practice which can be adapted and combined into a hybrid approach.

\section{Probing Collective Aesthetic Experience with Generative AI}

Generative AI art systems, whether prompt-based, multimodal, or embedded within creative software, operate through large-scale machine-learning architectures trained on massive corpora of images, text, videos, and other cultural artifacts. These training datasets do not simply serve as technical inputs; they effectively \emph{encode a collective aesthetic experience} accumulated from billions of user-generated materials circulating across social platforms, online archi\-ves, and digital repositories. As Manovich (2018) notes, ``gathered and aggregated data about the cultural behaviors of multitudes is used to model our ‘aesthetic self,’ predicting our future aesthetic decisions and tastes — potentially guiding us towards choices preferred by the majority.” \cite{manovich2018ai} In this sense, generative AI systems function as aesthetic modeling infrastructures that compute normative visual and stylistic conventions, reinforcing and amplifying the statistical patterns already prevalent within collective culture.

The images produced by these systems frequently reflect this inheritance. Because the models learn by identifying recurring structures, styles, and motifs in their training data, their outputs often exhibit extensive pattern repetition, recombination, and sophisticated mimicry. Practices such as neural style transfer exemplify how generative AI reproduces existing aesthetic modes, prompting critiques that much AI-generated art is an ``imitation work,” constrained by the patterns and biases embedded in its training material \cite{gatys2016style}. At the same time, the creation of the datasets themselves constitutes a profound socio-political intervention: the processes of data scraping, selection, labeling, and categorization define what the system can ``see,” establishing epistemic boundaries that shape how cultural materials are classified, normalized, and ultimately re-imagined. This transformation of diverse cultural expressions into fixed taxonomies is, in effect, an act of world-making encoded directly into the model’s structure.

Thus, the aesthetic output of generative AI is best understood not as autonomous creativity but as the materialization of statistical regularities across massive digital collections of human expression. These technologies mediate between individual intention and collective cultural memory, producing images that are simultaneously shaped by user agency and conditioned by the aggregated behaviors, preferences, and biases of the multitude. In other words, these AI systems act as a database of collective aesthetic experiences, which we can extract by providing appropriate contextual information.

\section{Text-Prompting as Style Synthesis Technique}

Generative art has been substantially reshaped by the emergence of diffusion-based image synthesis models, which now constitute the dominant paradigm for producing high-fidelity and stylistically diverse images \cite{rombach2022ldm}. Within these systems, text-prompting functions as the primary interface between human intention and computational generation. Text-to-image models translate linguistic descriptions into visual outputs, positioning language itself as a generative artistic material and reframing authorship around conceptual articulation rather than manual image production.

Platforms such as Midjourney, Stable Diffusion, and DALL-E exemplify this workflow and have been widely adopted across art and design practices. These systems enable users to invoke, recombine, and reinterpret historical, modern, and speculative visual styles with minimal technical barriers. As a result, aesthetic exploration that once required specialized expertise or labor-intensive workflows can now be conducted through iterative textual experimentation. In fields including digital art, product design, fashion, and architecture, text-to-image models increasingly function as tools for conceptual ideation, rapid prototyping, and speculative visualization rather than as final image generators \cite{liu2022prompt}.

Beyond efficiency and accessibility, text-prompting introduces a distinct mode of artistic empowerment. Artists are no longer confined to a single medium or stylistic lineage but can fluidly traverse and hybridize visual traditions through language. This capacity supports reinterpretation and critique, allowing practitioners to interrogate dominant aesthetic canons and speculate on emergent visual languages shaped by algorithmic processes \cite{manovich2018ai}. In this sense, text-to-image models operate less as autonomous creators and more as aesthetic mediators, extending human imagination through computational systems trained on large-scale cultural data. 

The efficacy of text-prompting depends strongly on the construction of the prompt itself. The emerging practice of promptology frames prompt design as a creative and epistemic act, emphasizing how textual structure guides both system behavior and user interaction \cite{liu2022prompt}. Prompts often follow an implicit compositional grammar, specifying style or historical reference, detailing subject matter and visual attributes, and concluding with affective cues indicating mood or conceptual tone. Through this layered structure, the prompt functions simultaneously as instruction, speculation, and site of aesthetic negotiation.

\section{Data Visualization and Art}
Information visualization has historically developed within a functionalist paradigm centered on perceptual efficiency, clarity, and analytical rigor. Jacques Bertin formalized visualization as a ``monose\-mic” sign system in which visual variables, such as position, size, value, color, orientation, and shape, are designed to communicate unambiguous meaning \cite{bertin1983semiology}. This rationalized grammar strongly influenced later approaches, most notably Edward Tufte’s advocacy for graphical integrity and minimalism, where aesthetic quality is understood to emerge from clarity and informational density rather than ornamentation \cite{tufte2001visual}. Cognitive and perceptual research further reinforced this orientation, framing visualization as an epistemic tool optimized for analytical tasks and correct interpretation \cite{ware1993information}.

However, this efficiency-driven model does not capture the full historical or cultural scope of visualization. Many visualization archetypes, such as trees, circles, and cosmograms, originate in symbolic, metaphysical, and philosophical diagrammatic traditions rather than scientific measurement \cite{lima2014trees}. Humanistic critiques emphasize that visualization is never neutral but always interpretive and situated, shaped by cultural assumptions and epistemological commitments \cite{drucker2011humanities}. These perspectives expand visualization beyond technical optimization toward aesthetic, rhetorical, and cultural interpretation.

Within this expanded framework, artistic data visualization emerges as a practice that treats data as expressive material rather than neutral information. Artistic visualization encourages contextualized reading, inviting viewers to reflect on social, cultural, or conceptual meanings instead of extracting singular analytical conclusions \cite{viegas2007artistic}. Central to this shift is the role of design. Research on information aesthetics and design in visualization demonstrates that visual forms, e. g., style, color, composition, and rhythm, are not decorative but constitutive of meaning, shaping how data is perceived, interpreted, and emotionally engaged \cite{lau2011design}. Visualization thus becomes an aesthetic and communicative act rather than a transparent analytical interface.

This perspective aligns with approaches that preserve or foreground the semantic and material qualities of data rather than reducing them to abstract graphical primitives \cite{lau2011information, manovich2010Data}. Artistic works that translate images, or sound into visual structures demonstrate how visualization can operate as an expressive language, privileging atmosphere, pattern, and perception over explanation.

Quantum mechanics provides a particularly rich foundation for generating data suitable for such aesthetic exploration. Mathematical formalisms such as the Schr\"{o}dinger equation or Hamiltonian eigenvalue problems produce complex numerical structures—proba\-bility distributions, wavefunctions, spectral patterns, entanglement measures, and temporally evolving interference phenomena. These quantum-generated datasets exhibit indeterminacy, non-classical geometry, and emergent structure, lending themselves naturally to artistic visualization. When visualized, they function simultaneously as scientific simulation and aesthetic speculation, bridging physical theory and generative art.

Within this context, data visualization can be understood as art when its intent shifts from efficient information transfer to conceptual and aesthetic inquiry; when visual form contributes directly to meaning through information aesthetics \cite{lau2011design}; when it embraces polysemy and interpretive openness rather than singular analytical conclusions \cite{viegas2007artistic} \cite{chen2014visualization}; and when it reflects critically on the constructed nature of data and representation itself \cite{drucker2011humanities}. Seen through this lens, quantum-generated data visualization represents a compelling frontier in contemporary media art, where scientific abstraction and aesthetic experience co-evolve.

\section{Top-Down Approach: Prompting with Generative AI}
\subsection{Text-Prompting}
This project employs Midjourney as the primary generative AI system for synthesizing visual outputs rendered in what we describe as a \emph{quantum style}. As a widely adopted text-prompt-based tool in contemporary design and artistic practice, Midjourney provides an accessible platform for examining how linguistic constructions translate into visual outcomes. In this study, generative control is exercised through structured prompt syntaxes rather than through any analysis of scientific quantum properties, such as superposition or entanglement. Instead, our focus is exclusively on the visual effects produced by prompt structure and parametric variation.

A central prompt formulation used throughout the project is:
\begin{equation}
    \texttt{quantum <scene>}
\label{eq:qs}
\end{equation}
where \texttt{<scene>} functions as a variable placeholder. The term \texttt{<scene>} is used broadly and can be replaced by the name of any artefact, space, place, environment, or scene-level configuration, such as a room, building, landscape, object-like artefact, or hybrid spatial setting.

\subsection{Style Synthesis with Weight-Adjustment}
\begin{figure}[H]
    \centering
    \includegraphics[width=\linewidth]{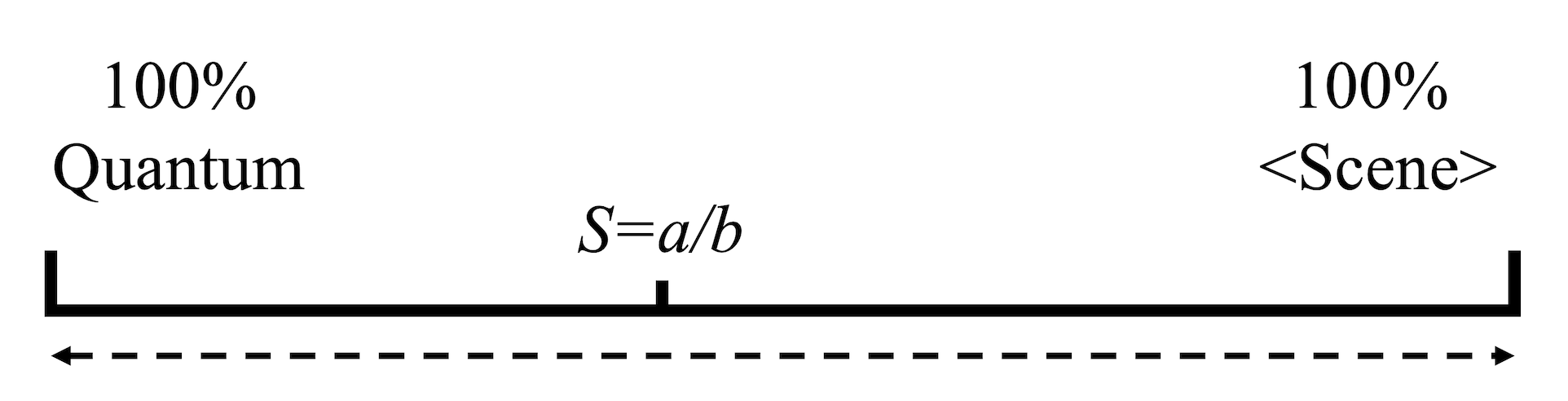}
    \caption{The synthesized image spectrum across pure quantum and scene images with the weighting ratio $S$.}
    \label{fig:1}
\end{figure}

Midjourney further supports a prompt-weighting mechanism that enables control over the relative influence of different textual components. This mechanism uses the syntax \texttt{``A::a B::b”}, in which \texttt{a} and \texttt{b} specify the proportional weights assigned to phrases \texttt{A} and \texttt{B}. We apply this feature to construct the formulation for text prompt:
\begin{equation}
    \texttt{quantum::a <scene>::b}
\label{eq:qs2}
\end{equation}
By varying the ratio $S=a/b$, we generate a continuous spectrum of images in which the visual characteristics of the generated scene shift between and increasingly abstract or transformed interpretations associated with the term \texttt{quantum} to one end, and recognizable representations of \texttt{<scene>} to the other, as illustrated in Figure \ref{fig:1}. This method allows for a tunable synthesis in which the contribution of \texttt{quantum} and \texttt{<scene>} can be systematically adjusted, producing families of related images rather than isolated outputs.

An important component of this workflow is the use of \emph{raw mode}, which minimizes Midjourney’s default aesthetic stylization. Raw mode is a generation setting that suppresses the system’s built-in preferences for cinematic lighting, color grading, and compositional embellishment, resulting in images that more closely reflect the textual prompt (available for version 5.1 or later). This reduced stylization is essential for isolating the visual effects produced by changes in prompt structure and weighting, ensuring that variations observed across outputs arise primarily from the linguistic parameters rather than from model-embedded aesthetic biases.

\begin{figure*}
    \centering
    \includegraphics[width=0.35\linewidth]{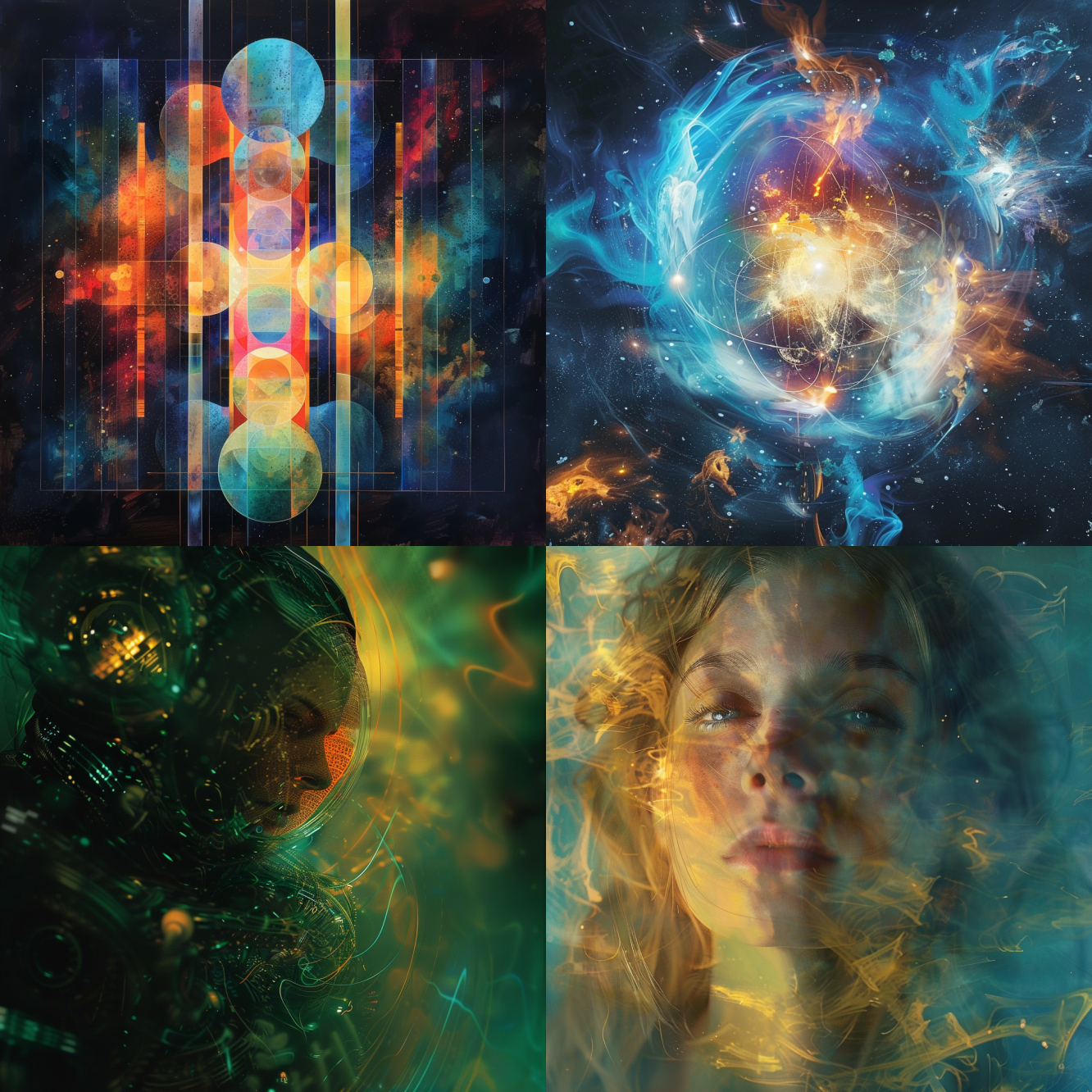}
    \hspace{0.02\linewidth}
    \includegraphics[width=0.35\linewidth]{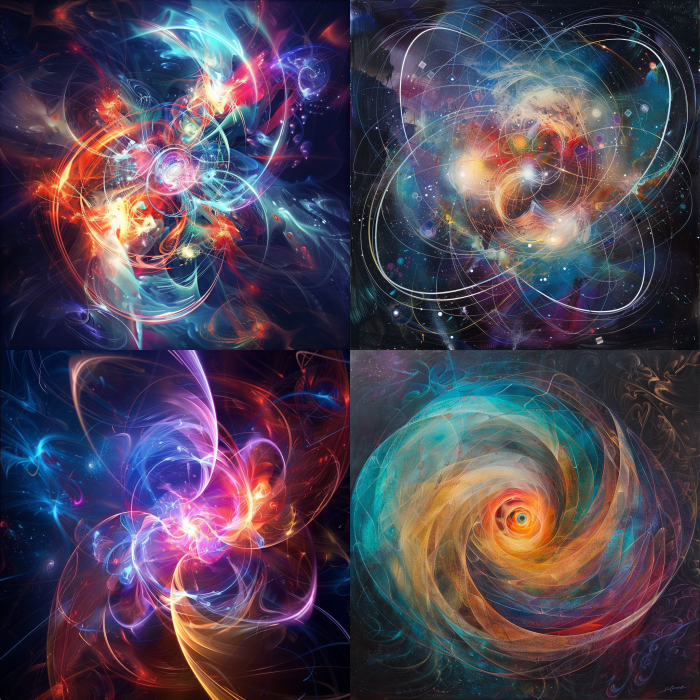}
    \caption{AI generated images using prompts ``\texttt{quantum}” (left) and ``\texttt{quantum {-}{-}style raw}” (right).}
    \label{fig:2}
\end{figure*}

\begin{figure*}
    \centering
    \includegraphics[width=0.35\linewidth]{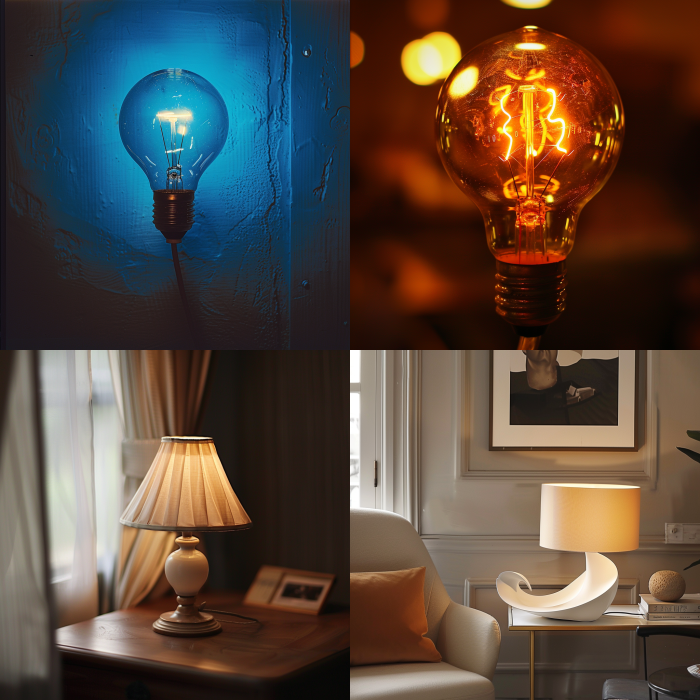}
    \hspace{0.02\linewidth}
    \includegraphics[width=0.35\linewidth]{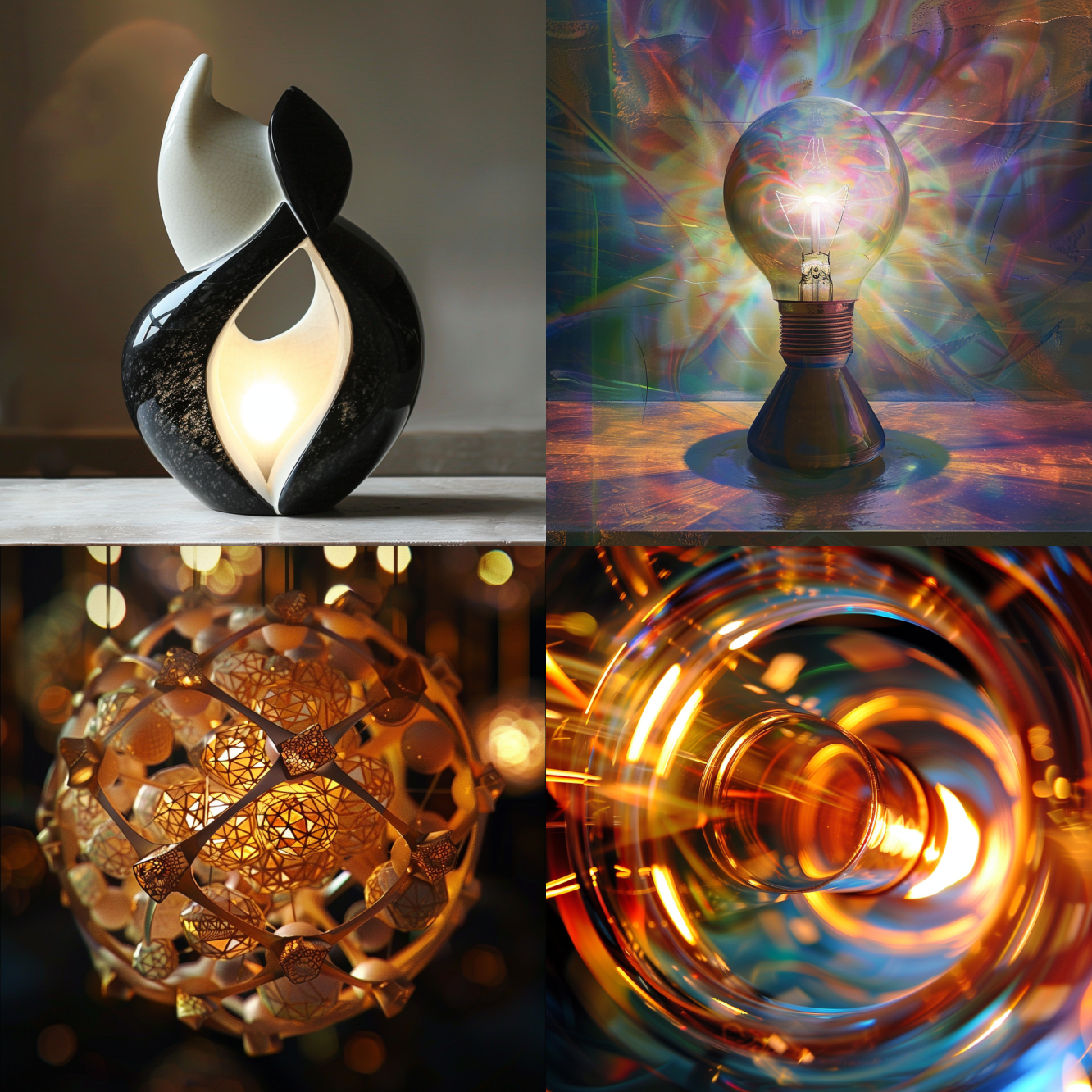}
    \caption{AI generated images using prompts ``\texttt{lamp {-}{-}style raw}” (left) and ``\texttt{quantum lamp}” with $S_B=0.5$ (right).}
    \label{fig:3}
\end{figure*}

\begin{figure*}
    \centering
    \includegraphics[width=0.35\linewidth]{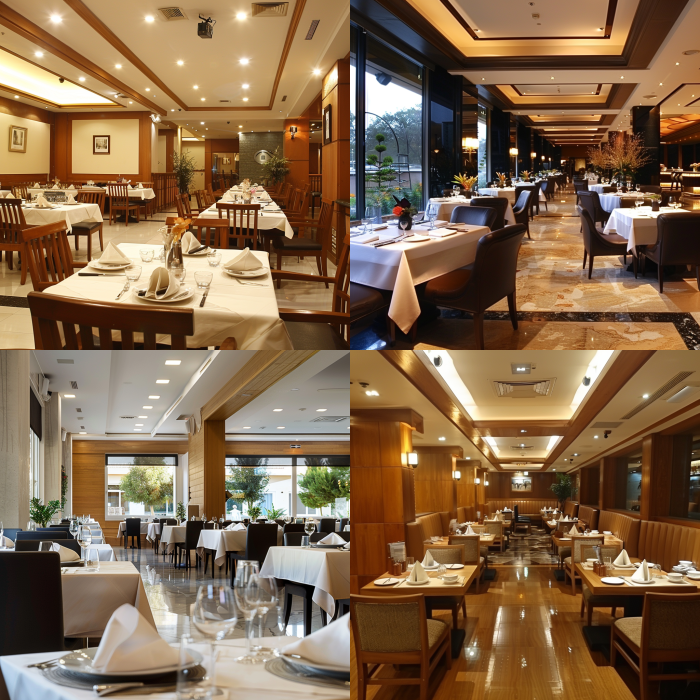}
    \hspace{0.02\linewidth}
    \includegraphics[width=0.35\linewidth]{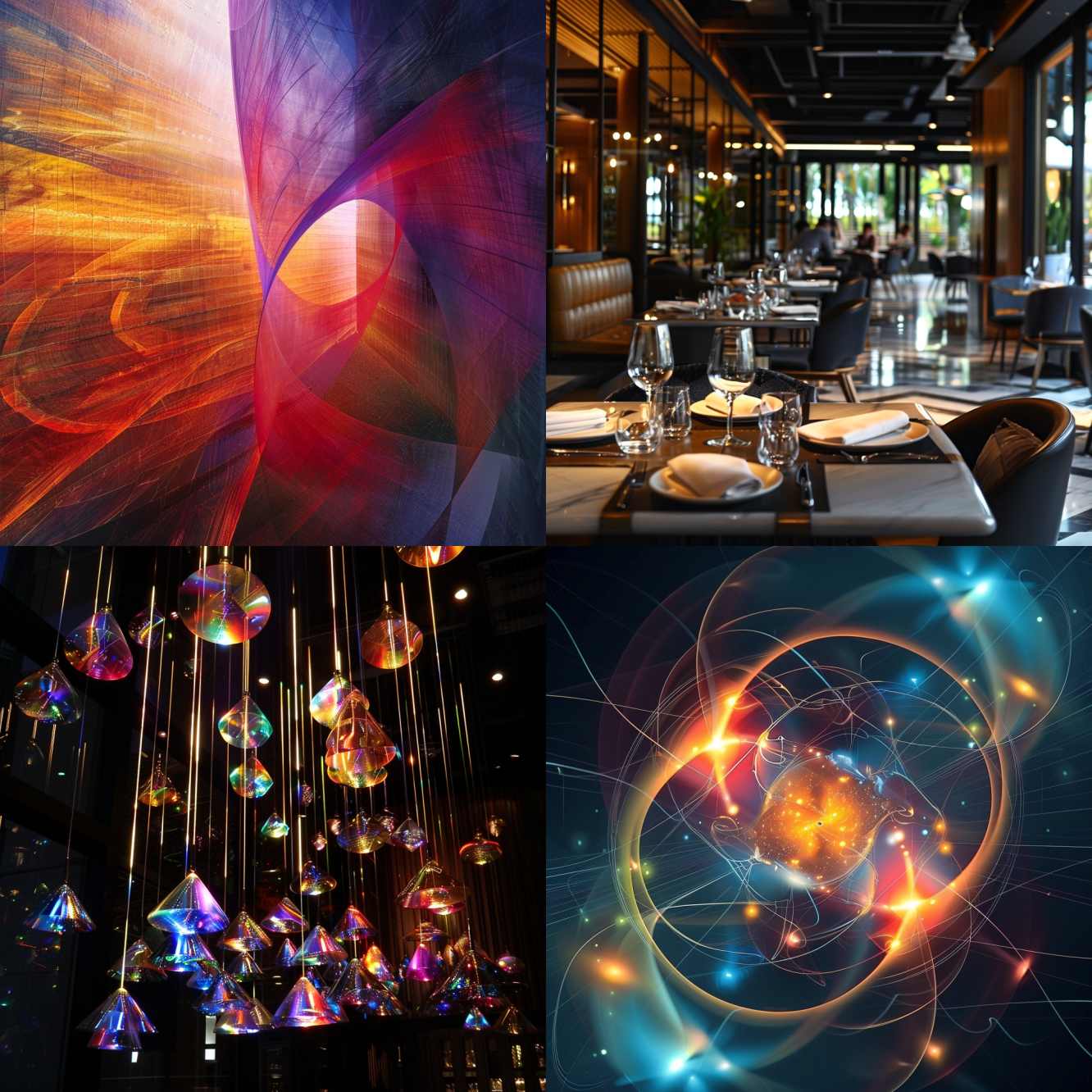}
    \caption{AI generated images using prompt ``\texttt{restaurant {-}{-}style raw}” and ``\texttt{quantum restaurant}” ratio $S_B=2.2$.}
    \label{fig:4}
\end{figure*}

In practice, the full prompt syntax used in this study combines both weighting and raw mode:
\begin{equation}
    \texttt{quantum::a <scene>::b {-}{-}style raw}
\label{eq:qs3}
\end{equation}

This configuration provides parametric control through the adjustable weight ratio $S$ and stylistic neutrality through the \texttt{{-}{-}style raw} option. Together, these features enable a controlled exploration of how the term ``quantum” visually reshapes different instantiated scenes, regardless of whether ``scene” refers to an artefact, a spatial environment, or a specific place. The resulting set of images forms a coherent and analyzable space in which quantum style is examined as an aesthetic effect produced through prompt-driven synthesis.

Furthermore, we assume that there exists a \emph{balance} ratio $S_B$, where the synthesis between the visual elements of \texttt{quantum} and \texttt{<scene>} are the most balanced. Away from this point of optimal blend, the synthesized image is either too much \texttt{quantum} or too much \texttt{<scene>}. In the current research, the judgement of this optimal blend relies on the researchers, which suffices for the present exploratory study. In the future, we plan to perform this judgement test by survey to provide a more objective judgement.

Through this workflow, Midjourney functions not only as a generative tool but also as an experimental environment. The prompt syntax \eqref{eq:qs3} establishes a repeatable method for investigating how linguistic modulation influences visual outcomes across a wide range of scene instantiations, supporting the articulation of quantum style as a speculative and generative aesthetic.

Here, to demonstrate the idea of quantum style synthesis by prompting, first we look at the visual of the prompt \texttt{quantum} in the raw mode. Figure \ref{fig:2} shows the difference between image generation with the raw mode on and off. Each time, Midjourney generates a set of four images. As can be seen in Figure \ref{fig:2}, the results without raw mode show often images with other distinct recognizable elements, such as person’s face, silhouette or geometrical shapes. By comparison, the generated images in raw mode show abstract figures with fluidity, florescent colors and galactic features. The latter thus are identified as the quantum visual features from the collective aesthetic experience.

Likewise, for the other end of the spectrum, namely, the scene, ``\texttt{<scene> {-}{-}style raw}” was used to generate its visual image. Figure \ref{fig:3} shows the generated image when \texttt{<scene>} is substituted with \texttt{lamp}. As can be seen, the four images largely conform to common visual expectations of lamps, albeit, the lower right one is more modern than the lower left, which is more of a classic style.

Using the prompt syntax \eqref{eq:qs3} for lamp, and scanning across the value of $S$, we found the balance ratio to be  $S_B=0.5$ , shown in Figure \ref{fig:3}. The top-left image presents a sculptural lamp with an internal glow emerging from a hollow core, suggesting quantum dualities of presence and absence, solidity and void, where light appears as an intrinsic state of the object. The top-right image reimagines a familiar light bulb as a source of prismatic dispersion, evoking photon emission and wave-particle ambiguity as light dissolves into a spectral field. The bottom-left image depicts a spherical lattice containing multiple glowing nodes, recalling atomic or molecular structures and suggesting quantized energy units within an ordered system. The bottom-right image abstracts the lamp into a swirling tunnel of light and reflection, where motion, chromatic distortion, and instability evoke continuous energy flow and uncertainty.

The procedure was repeated for ``restaurant” as the scene. Figure \ref{fig:4} shows the generated images for restaurant in raw mode. The balance ratio for synthesized ``quantum restaurant” was found to be $S_B=2.2$, shown in Figure \ref{fig:4}. The top-left image uses flowing, intersecting color fields to suggest an underlying energetic atmosphere, evoking quantum superposition through overlapping translucent planes. Although it contains fewer overt quantum elements, the top-right image presents a refined contemporary restaurant, where depth, reflections, and repetition create a calm yet complex spatial order, hinting at hidden structures beneath visual stability. The bottom-left image features suspended, iridescent glass forms that refract light into shifting spectra, transforming the dining space into an immersive field of luminous ``energy droplets.” The bottom-right image shows abstract swirling lines and glowing nodes orbiting a bright core, directly referencing atomic and orbital imagery, with little or no restaurant elements.

At the point of optimal blend, in both cases of ``quantum lamp” and ``quantum restaurant” the four images generated by Midjourney contain a mixture of balanced and unbalanced images, suggesting that this point indeed marks a transition point from a more quantum-weighted regime to a more scene-weighted one.

\begin{figure*}
    \centering
    \includegraphics[width=\linewidth]{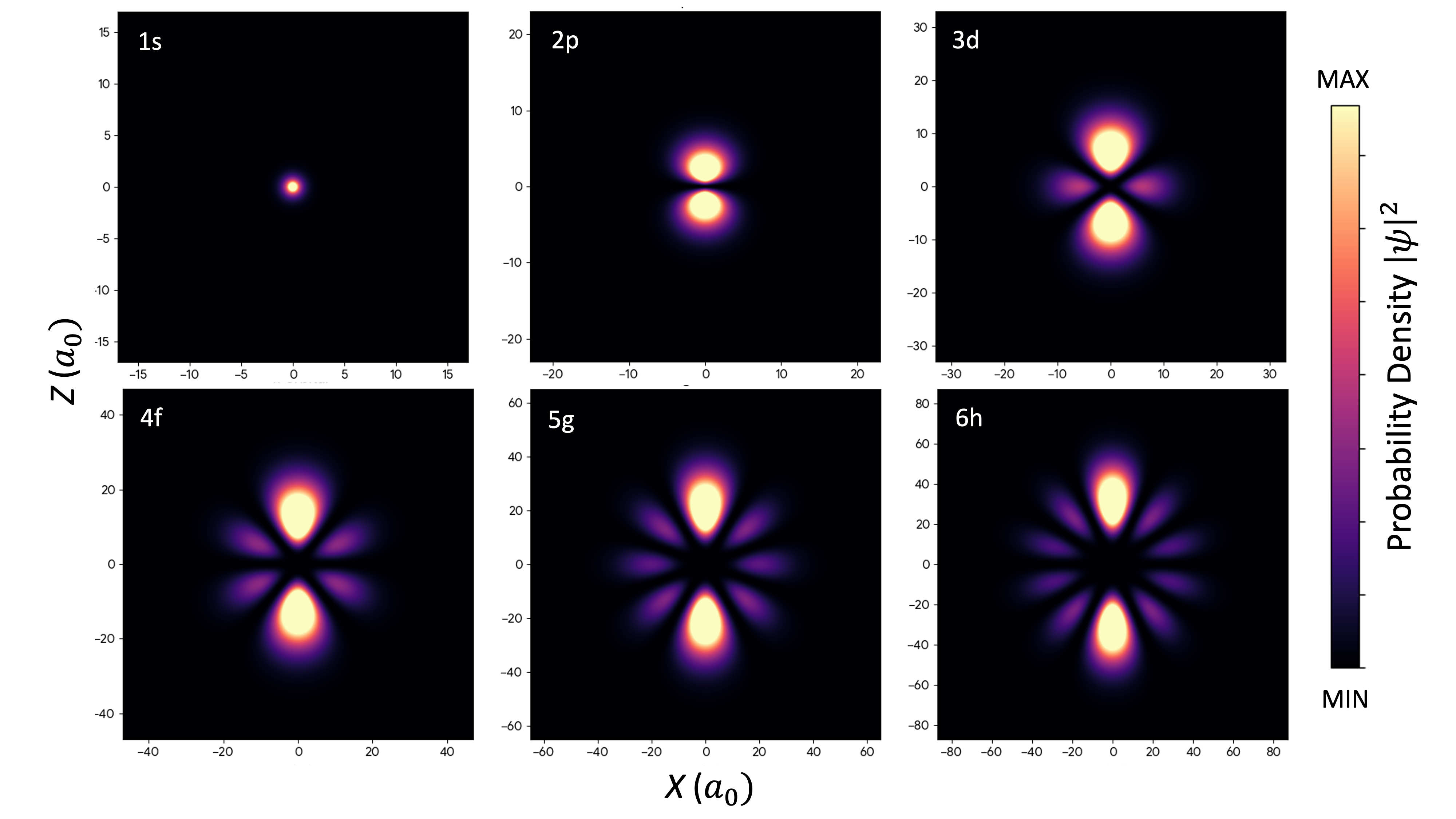}
    \caption{Visualization of the two-dimensional electron probability density of hydrogen wavefunction calculated from Equation~\eqref{eq:sch2}}
    \label{fig:5}
\end{figure*}

\section{Bottom-up Approach: Visualization from Quantum Data}
\subsection{Calculation of Atomic Orbitals}
In this approach we generate data from equations of quantum mechanics. The most fundamental equation in quantum mechanics is the general \emph{time-dependent} Schr\"{o}dinger equation:
\begin{equation}
    i\hbar \frac{d}{dt} |\psi(t)\rangle = \hat{H} |\psi(t)\rangle
\label{eq:sch}
\end{equation}
where $t$ is time, $|\psi(t)\rangle$ is the state vector of the quantum system, and $\hat{H}$ is the Hamiltonian operator, an observable of the system. To calculate observables of the system, $\hat{H}$ is substituted with an explicit mathematical form corresponding to the observable being calculated, e.g., energy or position.

Here, to illustrate the possible visualization outcome from the Schr\"{o}dinger equation, we take the case of hydrogen atom as an example, and calculate its \emph{time-independent} wavefunction, which can be derived from the time-dependent Equation \eqref{eq:sch} \cite{griffiths2005quantum}. The procedure yields a wavefunction for hydrogen atom as
\begin{equation}
    \resizebox{0.9\columnwidth}{!}{$
    \psi_{nlm}(r, \theta, \phi)
    =
    \sqrt{
    \left(\frac{2}{na_0}\right)^3
    \frac{(n-l-1)!}{2n[(n+l)!]}
    }
    \, e^\frac{-r}{na_0}
    \left(\frac{2r}{na_0}\right)^l
    L_{n-l-1}^{2l+1}\!\left(\frac{2r}{na_0}\right)
    Y_l^m(\theta,\phi)
    $}
\label{eq:sch2}
\end{equation}
where $(r, \theta, \phi)$ are the coordinates of the electron; $(n,l,m)$ are the quantum numbers, and $a_0$ is the Bohr radius.

The wavefunctions were calculated for atomic orbitals \emph{1s}, \emph{2p}, \emph{3d}, \emph{4f}, \emph{5g}, and \emph{6h}, corresponding to the quantum numbers $n=1~6$ and $l=0~5$, while $m=0$. The results of the calculation are visualized as two-dimensional probability density plots, $|\psi|^2$, in Figure \ref{fig:5}.

\subsection{Quantum Orbital as Artwork}
The six visualizations presented in Figure \ref{fig:5} depict hydrogen atomic orbitals rendered as two-dimensional cross-sections of the probability density $|\psi|^2$. Although grounded in quantum physics, these images can also be understood as aesthetic artifacts when examined through theories of information aesthetics and contemporary visualization discourse.

Formally, the series demonstrates a systematic progression of spatial complexity governed by quantum numbers. In Figure \ref{fig:5}, the \emph{1s} orbital exhibits spherical symmetry and a single probability maximum, resulting in a visually minimal and isotropic form. The \emph{2p} orbital introduces axial orientation and a nodal plane, dividing the distribution into two opposing lobes. Higher angular momentum states (\emph{3d}, \emph{4f}, \emph{5g}, and \emph{6h}) generate increasingly intricate geometries composed of multiple lobes and intersecting nodal surfaces. These nodal regions, where the wavefunction vanishes, are structurally constitutive rather than incidental. Visually, they function as negative space, shaping rhythm, symmetry, and balance through absence.

From the perspective of information aesthetics, the aesthetic quality of these images arises from the structured organization of information rather than expressive intent. Abstract mathematical descriptions are translated into perceptually legible patterns, revealing latent order embedded within the quantum system. As complexity increases across the series, coherence is maintained through symmetry and physical constraint, producing a balance between order and variation. This relationship resonates with Birkhoff’s aesthetic measure, in which aesthetic value emerges from the ratio between order and complexity \cite{birkhoff1933aesthetic}.

Within visualization theory, these orbital renderings exemplify how visual form functions as an epistemic medium rather than a neutral illustration. Continuous color gradients map scalar probability values to luminance and hue, enabling intuitive perception of gradients, symmetries, and nodal hierarchies. Such mappings foreground relational structure and pattern, supporting arguments that design plays a central role in transforming data into meaningful and aesthetically engaging visual experiences \cite{lau2011information}.

At the perceptual level, the images engage viewers both cognitively and affectively. Recognition of symmetry and nodal structure supports analytical reading, while the luminous, flower-like geometries evoke contemplation and wonder. Accordingly, these hydrogen orbital visualizations may be understood as information aesthetic artworks: visual manifestations of quantum structure that operate simultaneously as scientific representations and aesthetic objects, demonstrating how visualization can become a site where knowledge, perception, and aesthetic experience converge.

\section{Conclusion}
This study has proposed a dual methodological framework for approaching quantum aesthetics, combining top-down generative AI practices with bottom-up visualization derived from quantum-mechanical data. Rather than attempting to resolve the elusive nature of quantum aesthetics into a single visual language, the roadmap presented here foregrounds plurality, mediation, and emergence as defining characteristics. Across both approaches, quantum aesthetics appear not as direct representations of physical reality, but as negotiated constructs shaped by linguistic framing, computational systems, cultural memory, and mathematical constraint.

Within this roadmap, the top-down approach represents a pioneering direction. By appending quantum to familiar artefacts, spaces, or environments, the prompt destabilizes everyday visual expectations and reimagines the scene through metaphors of energy, indeterminacy, and transformation. It systematically employs text-prompt-based generative AI as an epistemic instrument for probing quantum aesthetics as a collective cultural construct. By treating ``quantum” as a linguistically and culturally charged modifier, this approach maps how quantum ideas are already internalized, aestheticized, and circulated within contemporary visual culture. Prompt weighting and controlled synthesis function as navigational tools along this path, revealing transitional thresholds between recognizability and abstraction. In this sense, the top-down route charts how quantum aesthetics emerge through cultural imagination as encoded in large-scale generative models, rather than through correspondence with physical theory.

The bottom-up approach delineates a contrasting route along the roadmap, beginning from quantum mechanics itself. In this paper, the calculation and visualization of hydrogen atomic orbitals serve as a concrete example of how aesthetic form can emerge directly from physical law under strict mathematical constraint. However, this example represents only one possible entry point. Other quantum-mechanical equations, computational systems, numerical simulations, or experimentally measured quantum data may likewise serve as generative sources for artistic visualization. Along this path, quantum aesthetics arise from structured complexity, symmetry, indeterminacy, and the interplay between presence and absence inherent in physical systems, independent of stylistic or cultural intention.

These two approaches mark orthogonal yet interconnected directions within the proposed roadmap. The top-down path operates through language, cultural abstraction, and collective imagination, while the bottom-up path operates through physical law, data generation, and material constraint. Crucially, the roadmap does not prescribe these paths as separate or exclusive. They may be adapted, intersected, and combined into hybrid approaches in which quantum data informs generative systems, or generative AI is used to speculate on, reinterpret, or recontextualize physically grounded quantum structures. Such hybrid practices point toward new territories where cultural imaginaries and physical realities co-evolve.

In this proposed framework, quantum aesthetics is treated as a navigable field rather than a fixed destination. This roadmap thus positions artistic practice as a mode of epistemic exploration—one that does not merely communicate quantum knowledge, but actively reshapes how quantum phenomena are perceived, felt, and imagined. It offers fertile ground for future artistic experimentation as well as pedagogical applications, where generative tools and quantum data can be integrated into learning environments to foster experiential understanding, critical reflection, and speculative engagement with quantum science.

\section*{Acknowledgements}
The project was partially funded by National Science and Technology Council Grant No. 114-2420-H-A49-001 and 111-2410-H-A49-003-MY2.

\bibliographystyle{unsrt}
\bibliography{refs2}

@article{putnam2015hydrogen,
  title={Studies in Composing Hydrogen Atom Wavefunctions},
  author={Putnam, L. J. and Kuchera-Morin, JoAnn and Peliti, L.},
  journal={Leonardo},
  volume={48},
  number={2},
  pages={158--166},
  year={2015}
}

@article{kuchera2022complex,
  title={Composing and Performing Complex Systems: From the Quantum to the Cosmological},
  author={Kuchera-Morin, JoAnn A.},
  journal={Leonardo},
  volume={55},
  number={1},
  pages={30--33},
  year={2022}
}

@book{thomas2018quantum,
  title={Quantum Art and Uncertainty},
  author={Thomas, Paul},
  publisher={Intellect},
  address={Bristol},
  year={2018}
}

@article{heaney2019quantum,
  title={Quantum Computing and Complexity in Art},
  author={Heaney, Libby},
  journal={Leonardo},
  volume={52},
  number={3},
  pages={230--235},
  year={2019}
}

@misc{las2020sensing,
  title={Sensing Quantum},
  author={{LAS Art Foundation}},
  address={Berlin},
}

@incollection{manovich2010Data,
  author    = {Manovich, Lev},
  title     = {Data Visualization as New Media Aesthetics},
  booktitle = {Visualizing Culture},
  editor    = {Drucker, Johanna and others},
  publisher = {MIT Press},
  address   = {Cambridge, MA},
  year      = {2010}
}

@book{manovich2018ai,
  title={AI Aesthetics},
  author={Manovich, Lev},
  publisher={Strelka Press},
  year={2018}
}

@article{lau2011information,
  title={Towards a Model of Information Aesthetics in Information Visualization},
  author={Lau, Allen Y. S. and Vande Moere, Andrew},
  journal={IEEE Computer Graphics and Applications},
  volume={31},
  number={3},
  pages={87--92},
  year={2011}
}

@inproceedings{gatys2016style,
  title={Image Style Transfer Using Convolutional Neural Networks},
  author={Gatys, Leon A. and Ecker, Alexander S. and Bethge, Matthias},
  booktitle={Proceedings of CVPR},
  pages={2414--2423},
  year={2016}
}

@inproceedings{rombach2022ldm,
  title={High-Resolution Image Synthesis with Latent Diffusion Models},
  author={Rombach, Robin and Blattmann, Andreas and Lorenz, Dominik and Esser, Patrick and Ommer, Bj{\"o}rn},
  booktitle={Proceedings of CVPR},
  pages={10684--10695},
  year={2022}
}

@inproceedings{liu2022prompt,
  title={Design Guidelines for Prompt Engineering Text-to-Image Generative Models},
  author={Liu, Vivian and Chilton, Lydia B.},
  booktitle={Proceedings of CHI},
  pages={Article 665},
  year={2022}
}

@book{bertin1983semiology,
  title={Semiology of Graphics: Diagrams, Networks, Maps},
  author={Bertin, Jacques},
  translator={Berg, William J.},
  publisher={University of Wisconsin Press},
  year={1983},
  note={Originally published 1967}
}

@book{tufte2001visual,
  title={The Visual Display of Quantitative Information},
  author={Tufte, Edward R.},
  edition={2},
  publisher={Graphics Press},
  year={2001}
}

@book{ware1993information,
  title={Information Visualization: Perception for Design},
  author={Ware, Colin},
  publisher={Morgan Kaufmann},
  address={San Francisco},
  year={1993}
}

@book{lima2014trees,
  title={The Book of Trees; The Book of Circles},
  author={Lima, Manuel},
  publisher={Princeton Architectural Press},
  year={2014-2017}
}

@article{drucker2011humanities,
  title={Humanities Approaches to Graphical Display},
  author={Drucker, Johanna},
  journal={Digital Humanities Quarterly},
  volume={5},
  number={1},
  year={2011}
}

@incollection{viegas2007artistic,
  title={Artistic Data Visualization: Beyond Visual Analytics},
  author={Vi{\'e}gas, Fernanda B. and Wattenberg, Martin},
  booktitle={Online Communities and Social Computing},
  editor={Schuler, Douglas},
  pages={182--191},
  publisher={Springer},
  year={2007}
}

@article{lau2011design,
  title={On the Role of Design in Information Visualization},
  author={Lau, Andrew and Vande Moere, Avner},
  journal={Information Visualization},
  volume={10},
  number={4},
  pages={356--371},
  year={2011}
}

@article{chen2014visualization,
  title={What Is Visualization Really For?},
  author={Chen, Min and Floridi, Luciano and Borgo, Rita},
  journal={IEEE Computer Graphics and Applications},
  volume={34},
  number={3},
  pages={12--17},
  year={2014}
}

@book{griffiths2005quantum,
  title={Introduction to Quantum Mechanics},
  author={Griffiths, David J.},
  edition={2},
  publisher={Pearson},
  year={2005}
}

@book{birkhoff1933aesthetic,
  title={Aesthetic Measure},
  author={Birkhoff, George David},
  publisher={Harvard University Press},
  year={1933}
}
\section*{Author Biographies}
Ivan Liu is an artist and a researcher. He studied physics at Imperial College London, and subsequently obtained his PhD from Max-Planck Institute for the Physics of Complex Systems in Germany. He worked as an independent artist for 10 years before becoming a full-time faculty member at the National Yang Ming Chiao Tung University (NYCU) in Taiwan, where he founded the Future Narratives Lab. His current works explore new ways to narrate contemporary issues with science-inspired artistic representations. His works have been exhibited in renowned venues such as Ars Electronica (AT), IRCAM (FR), Chroniques Biennales (FR), ISEA 2025 (KR), LEV Festival (ES), National Taiwan Museum of Fine Art (TW), and Kaohsiung Museum of Fine Art (TW).

H. Y. Chen is currently a MA student at the Future Narratives Lab, NYCU.
\end{document}